\documentclass[aps,prl,preprint,superscriptaddress]{revtex4-1}

\usepackage[english]{babel}
\usepackage{graphicx}
\usepackage{subfigure}  
\usepackage{bm}
\usepackage{amsmath} 
\usepackage{epstopdf}
\AppendGraphicsExtensions{.tif} 
\usepackage{epsfig} 
\usepackage{subfigure} 
\usepackage{fancyvrb} 
\usepackage[usenames,dvipsnames]{color} 
\usepackage[normalem]{ulem} 
\usepackage{tikz} 
\newcommand{\Df}[2]{\ensuremath{\frac{d #1}{d #2}}}

\begin{document}

\title{Active Thermal Extraction of Near-field Thermal Radiation}


\author{D. Ding }
\author{A. J. Minnich}%
 \email{aminnich@caltech.edu}
\affiliation{ 
Division of Engineering and Applied Science, California Institute of Technology, Pasadena, California 91125, USA}%

\date{\today}

\begin{abstract}
Radiative heat transport between materials supporting surface-phonon polaritons is greatly enhanced when the materials are placed at sub-wavelength separation as a result of the contribution of near-field surface modes. However, the enhancement is limited to small separations due to the evanescent decay of the surface waves. In this work, we propose and numerically demonstrate an active scheme to extract these modes to the far-field. Our approach exploits the monochromatic nature of near-field thermal radiation to drive a transition in a laser gain medium, which, when coupled with external optical pumping, allows the resonant surface mode to be emitted into the far-field. Our study demonstrates a new approach to manipulate thermal radiation that could find applications in thermal management. 
\end{abstract}


\maketitle

Thermal radiation plays a role in many applications ranging from infrared detection and sensing applications for environmental and medical studies \cite{kendall_applied_1966,werle_near-and_2002} to energy harvesting with solar thermophotovaltics \cite{swanson_proposed_1979,bermel_tailoring_2011,lenert_nanophotonic_2014} and infrared emissions from Earth to space \cite{byrnes_harvesting_2014}. Thermal radiation is also essential to thermal management applications as in microelectronics \cite{buller_effects_1988}, space technology \cite{jenness_radiative_1960} and buildings \cite{raman_passive_2014}. 

In the far-field, the blackbody limit governs the maximum radiative flux between two bodies. Recently, a number of works have demonstrated that near-field radiative heat transfer is enhanced by many orders of magnitude compared to the far-field limit for closely spaced objects with either natural \cite{shen_surface_2009,rousseau_radiative_2009} or engineered resonant surface modes \cite{rodriguez_frequency-selective_2011,biehs_hyperbolic_2012,guo_thermal_2013,miller_effectiveness_2014,shi_near-field_2015}. There have also been efforts to couple these near-field modes into the far-field with the use of grating structures \cite{greffet_coherent_2002}, antennas \cite{schuller_optical_2009}, and a thermal extraction lens \cite{yu_enhancing_2013}. However, all of the mentioned techniques to control radiation are passive and are still subject to the blackbody limit in the far-field. 

Unlike passive schemes, active schemes extract energy from a system through external work, allowing certain limits to be overcome without violating the second law of thermodymanics. For example, refrigerators move heat from cold to hot using electrical work input. In optics, external work in the form of laser light has been used to cool of gaseous matter to sub-millikelvin temperatures \cite{hansch_cooling_1975,phillips_nobel_1998} by removing kinetic energy from the atoms. In solid-state materials, optical irradiation can also cool materials by emission of upconverted fluorescence \cite{pringsheim_zwei_1929} due to removal of energy in the form of phonons. This concept has been experimentally demonstrated to cool rare-earth doped glass \cite{sheik-bahae_optical_2007,seletskiy_laser_2010} to cryogenic temperatures and recently to cool semiconductors by 40 K from the ambient temperature \cite{zhang_laser_2013}. However, no active schemes are available to extract energy out of a system as thermal radiation. 

Here, we theoretically propose and numerically demonstrate an active thermal extraction scheme that extracts near-field thermal photons into the far-field. Our approach exploits the monochromatic nature of near-field thermal radiation to drive a transition in a laser gain medium, which, when coupled with external optical pumping, allows the resonant surface mode to be emitted into the far-field. Unlike the passive case, our active approach can lead to thermal radiative flux exceeding the blackbody limit in the far-field because work is being performed on the system.

\begin{figure}[h!]
\centering
\includegraphics[scale=0.5]{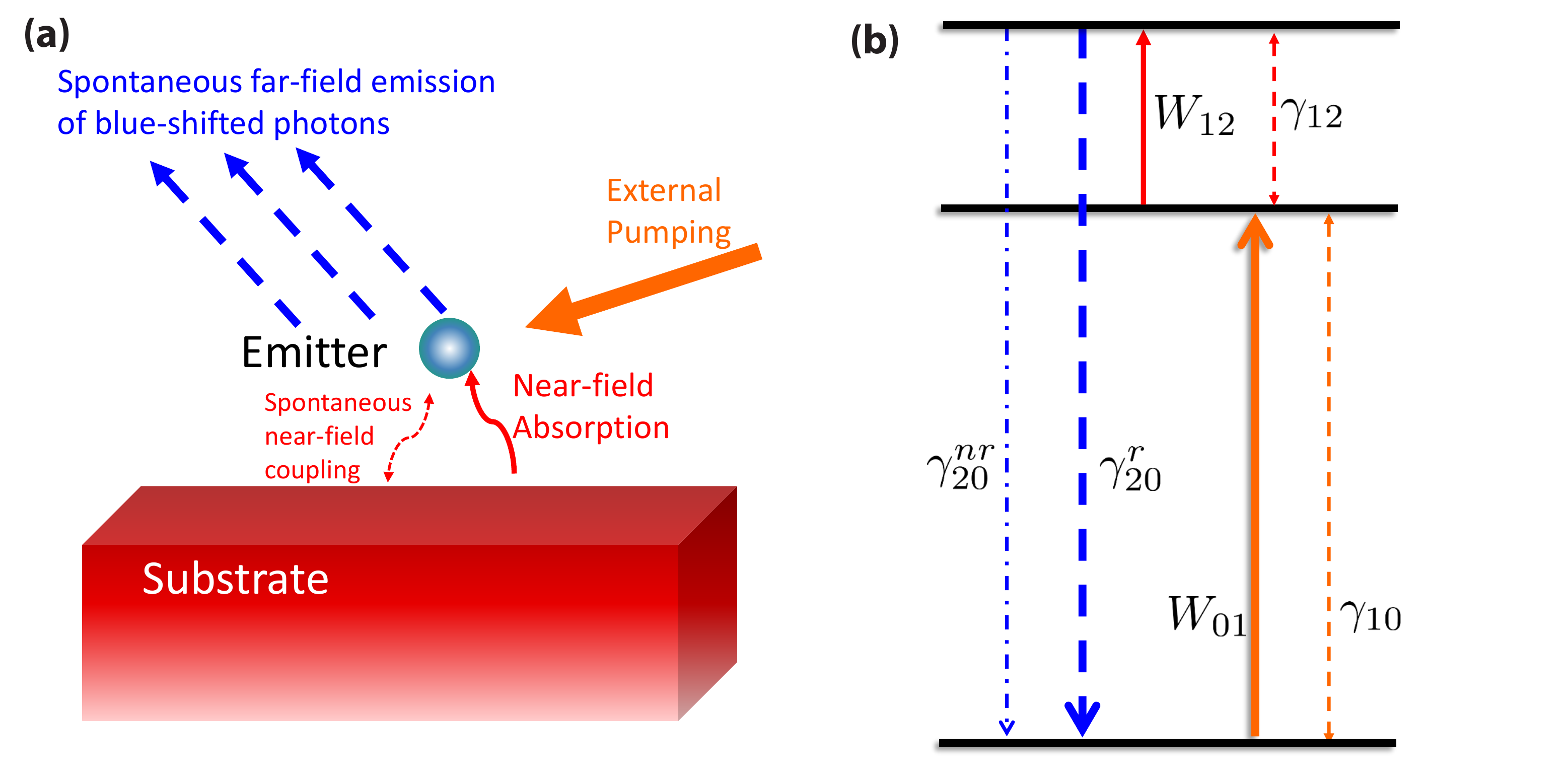}
\caption{\label{fig:schematic} (a) Schematic of the active thermal extraction scheme. An emitter with discrete energy levels is placed in the near-field region of a semi-infinite planar substrate supporting a surface resonance. The external pumping couples with the near-field energy to be emitted as blue-shifted spontaneous emission in the far-field. Various other spontaneous decay channels into the near-field and the far-field are also taken into account. (b) Energy level diagram of the emitter for our proposed concept. The 0-1 transition absorbs external pump photons, and near-field photons drive the 1-2 transition. Spontaneous emission from the 2-0 transition emits near-field photon to the far-field. $\gamma_{ij}$ are spontaneous decay rates for the i-j transition with the superscript "r" for radiative and "nr" for non-radiative rate. $W_{ij}$ are the absorption rate that depends on the energy density. The orange arrow indicates external optical pumping, the dashed arrows indicate various spontaneous decay channels with the blue arrows indicating the upconverted emitted photons carrying near-field energy into the far-field.}
\end{figure}

A schematic of the method is given in Fig.\,\ref{fig:schematic}(a). A laser gain medium containing emitters with discrete energy levels is placed in the near-field of a material that supports a resonant surface wave. We model the emitters as a three-level system, as shown in Fig.\,\ref{fig:schematic}(b). An external pump laser is tuned to the 0-1 transition, exciting population into level 1. If thermal radiation drives the transition from 1-2 and the 2-0 transition is radiative with high quantum efficiency, the electron transition will emit blue-shifted photons in the far-field, thereby extracting the trapped near-field thermal radiation.

With a typical blackbody spectrum, the efficiency of such a scheme would be vanishingly small because of the low energy density and the broadband nature of thermal radiation \cite{siegman_lasers_1986}. However, in the near-field, it has been demonstrated that the radiative energy density is nearly monochromatic and far exceeds that in the far-field by several orders of magnitude \cite{shchegrov_near-field_2000}. Therefore, with near-field thermal radiation the 1-2 transition can be efficiently driven by matching the near-field energy resonance energy to the 1-2 transition energy. 

To study this system, we use rate equations to determine the steady-state populations in each energy level with external and near-field pumping:
\begin{align}
\Df{N_2}{t} &= -W_{12}(N_2-N_1) - \gamma_{12}N_2 - \gamma_{20}N_2 \label{Eq:rate1}\\ 
\Df{N_1}{t} &= W_{12}(N_2-N_1) -W_{01}(N_1-N_0) - \gamma_{10}N_1 + \gamma_{12}N_2  \label{Eq:rate2}\\
\Df{N_0}{t} &= W_{01}(N_1-N_0) + \gamma_{20}N_2 + \gamma_{10}N_1 \label{Eq:rate3}\\ 
N_t &= N_0+N_1+N_2 \label{Eq:rate4}
\end{align}
where $W_{12}$ is the absorption rate of the 1-2 transition as a result of the near-field energy density, $W_{01}$ is the absorption rate of the 0-1 transition due to external pumping, $N_i$ are population density of each level, $N_t$ is the total population density for system and $\gamma_{ij}$ is the overall (radiative and non-radiative) spontaneous decay rate of the i-j transition. Here, $\gamma^r_{ij}$ stands for radiative rate of the i-j transition such that $\gamma_{ij}=\gamma^r_{ij}+\gamma^{nr}_{ij}$. We assume that all energy levels are non-degenerate so that $W_{ij}=W_{ji}$. Solving Eqs. \ref{Eq:rate1} to \ref{Eq:rate4} in steady state yields the equilibrium population densities for each level from which the power density can be expressed as
\begin{align}
P_{01}&=\hbar\omega_{10}W_{01}(N_0-N_1) \nonumber \\
&=\frac{\hbar\omega_{10} N_t W_{01} \left( W_{12}(\gamma_{10}+\gamma_{20})+\gamma_{10}(\gamma_{12}+\gamma_{20})\right)}{W_{12} (\gamma_{10} + \gamma_{20}) + \gamma_{10} (\gamma_{20} + \gamma_{12}) + 
 W_{01} \left(3 W_{12} + 2 (\gamma_{20} + \gamma_{12})\right)} \label{Eq:P10}\\ 
 P_{20,net} &= \hbar(\omega_{20}-\omega_{10})\gamma^{r}_{20}N_2 \nonumber\\
&= \frac{\hbar (\omega_{20}-\omega_{10}) N_t W_{01}\gamma^{r}_{20}W_{12}}{W_{12} (\gamma_{10} + \gamma_{20}) + \gamma_{10} (\gamma_{20} + \gamma_{12}) + 
 W_{01} \left(3 W_{12} + 2 (\gamma_{20} + \gamma_{12})\right)}
 \label{Eq:P20}
\end{align}
where $P_{01}$ is the external power density absorbed by the 0-1 transition and $P_{20,net}$ is the net extracted power density into the far-field from the 2-0 transition.

Using Eqs. \ref{Eq:P10} and \ref{Eq:P20}, the intrinsic efficiency of extraction can be expressed as the ratio of the amount of extracted energy radiated into the far-field by the 2-0 transition with respect to the external pump energy absorbed by the 0-1 transition
\begin{equation}
\eta_{10}=\frac{P_{20,net}}{P_{01}}=\frac{(\omega_{20}-\omega_{10}) \gamma^r_{20} W_{12}}{\omega_{10} (W_{12}(\gamma_{20}+\gamma_{10})+\gamma_{10}(\gamma_{20}+\gamma_{12}))}
\label{Eq:eff10}
\end{equation}

In the ideal limit of a dominant radiative 2-0 transition $\gamma_{20}$ and strong near-field absorption $W_{12}$, Eq. \ref{Eq:eff10} tends towards $(\omega_{20}/\omega_{10}-1)(\gamma^r_{20}/\gamma_{20})$ which depends intuitively on the ratio of the emitted net energy and absorbed photon energy and on the radiative rate of the 2-0 transition for the photons that reach the far-field. When $\eta_{10}>0$, there is net energy extracted from the system if we assume no parasitic absorption of external pump energy. The intrinsic efficiency in Eq. \ref{Eq:eff10} depends only on internal parameters of the system and is independent of the absorption rate $W_{01}$ of the external pumping (0-1) transition.   

To estimate the efficiency of the scheme, we take properties based on rare-earth dopant embedded in gallium lanthanum sulfide (GLS) chalcogenide glass as the emitter system with typical values listed in Table \ref{Tab:properties} \cite{schweizer_infrared_1999,seddon_progress_2010}. Such a system has a high quantum yield for transitions in the mid-infrared (MIR) region with a high melting point \cite{petrovich_near-ir_2003} and has been used in making fiber-based MIR lasers \cite{seddon_progress_2010}. Here, we assume the quantum efficiency of the equivalent 0-1 and 2-1 transitions to be unity and choose the wavelength-independent permittivity of the GLS chalcogenide glass to be 4.8 \cite{yayama_refractive_1998}.
\begin{table}[h]
\begin{center}
\begin{tabular}{c c c c c}
 \hline\hline
Transition & $\lambda (\mu$m) & $\gamma^0_{ij} (\text{s}^{-1})$ & QE (\%) \\
\hline
0-1 & 1.83 & 1034 & 100 \\
2-0 & 1.22 & 1370 & 79 \\
1-2 & 3.88 & 36 & 100 \\
\hline\hline
\end{tabular}
\caption{\label{Tab:properties} Parameters of a typical rare-earth emitter in GLS chalcogenide glass for modeling our proposed system. $\gamma^0_{ij} (\text{s}^{-1})$ stands for the decay rate of the i-j transition for an isolated emitter and QE is the quantum efficiency of the transition.} 
\end{center}
\end{table}
Then, we model the substrate permittivity with the expression $\epsilon(\omega)=\epsilon_{\infty}(\omega^2_L-\omega^2-i\gamma \omega)/(\omega_T^2-\omega^2-i\gamma\omega)$ where $\epsilon_{\infty}=5.3$, $\omega_T=388.4\times10^{12} \text{ s}^{-1}$,$\omega_L=559.3\times10^{12} \text{ s}^{-1}$ and $\gamma=0.9\times10^{12} \text{ s}^{-1}$. We tailor the substrate resonance to match the 1-2 transition with Re($\epsilon_{substrate}(\omega))=-\epsilon_{medium}$ so as to enhance the energy density of the near-field thermal radiation with the emitter. 
 
\begin{figure}[h!]
\includegraphics[scale=0.37]{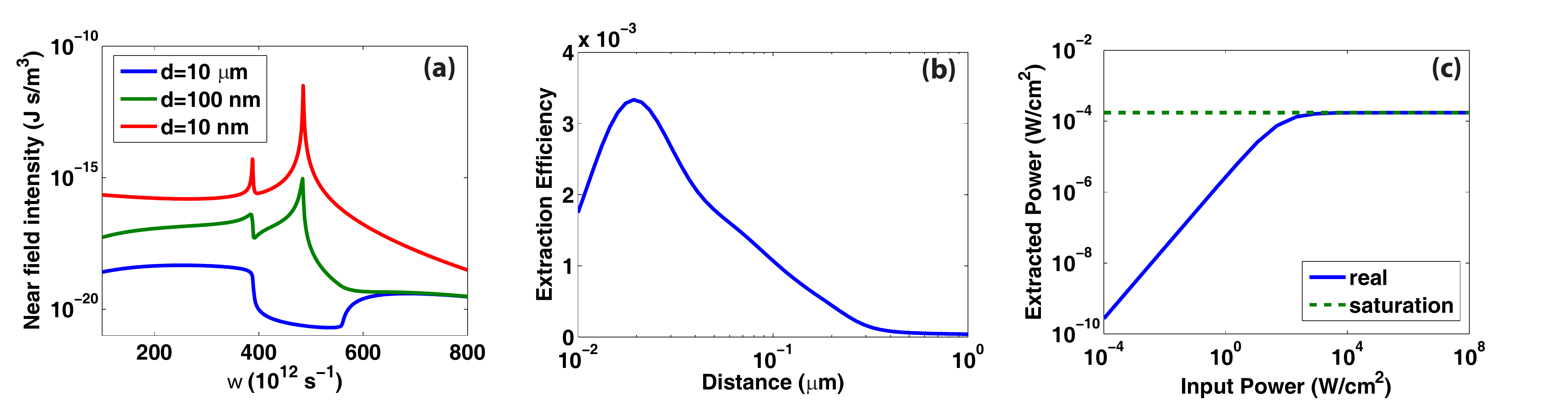}
\caption{\label{fig:eff_real} (a) Energy density at different distances $d$ from the surface of the substrate. The top medium is GLS chalcogenide glass. The thermal radiation spectrum becomes nearly monochromatic close to the substrate, consistent with other calculations \cite{shchegrov_near-field_2000}. (b) Extraction efficiency $\eta_{10}$ of external pumping from the 0-1 transition assuming properties in Table \ref{Tab:properties}. The low efficiency in the blue line is a result of the enhanced spontaneous rate for 1-2 transition in Fig.\,\ref{fig:gamma21}. (c) Integrated power extracted for emitters uniformly distributed from 10 nm onward from surface. The density of emitters is assumed to be $10^{20}\,\text{cm}^{-3}$. The saturation behavior approaches the green dashed "saturation" line due to the finite number of emitters in the system saturating the population difference at high input powers.}
\end{figure}

To calculate the intrinsic extraction efficiency of this system using Eq.\,\ref{Eq:eff10}, we need to know near-field absorption rate $W_{12}$, which in turn requires the near field energy of the substrate. To obtain this parameter, we use the formulation from Ref.\,\cite{joulain_surface_2005} to calculate the near-field energy density $I(\omega)$ of the substrate at a temperature of 750 K where the blackbody spectrum peaks around $3.9\,\mu$m, matching the 1-2 transition wavelength in Table \ref{Tab:properties}. Fig. \ref{fig:eff_real}(a) shows that the near-field energy density $I(\omega)$ increases by seven to eight orders of magnitude compared to the far-field energy density at very small distances from the substrate. Also, the spectrum peaks at $\omega=485.5\times10^{12} \text{ s}^{-1}$ when Re($\epsilon_{substrate}(\omega))=-\epsilon_{medium}$. Then, we can approximate the near-field absorption rate using of a recent formulation by Archambault et al. \cite{archambault_quantum_2010} which specifies the modified spontaneous and stimulated transition rates by quantizing a surface plasmon for the planar interface. The isotropic stimulated rate in Eq. (29) of Ref.\,\cite{archambault_quantum_2010} is also the absorption rate $W_{12}$ for the near-field in Eqs.\,\ref{Eq:rate1} to \ref{Eq:rate4}. We modify this expression to incorporate the energy per unit volume $I(\omega)=\int^{\infty}_{0}I(\omega,k) dk$ in Fig.\,\ref{fig:eff_real}(a) for the transition for different values of wave vector $k$ to obtain 
\begin{align}
W_{ij,near-field}&=\frac{\gamma^0_{ij} \pi^2 c^3}{2\hbar \omega_0^3}\int^{\infty}_{-\infty} \int^{\infty}_{0} (1+\lvert\frac{k}{\sqrt{\epsilon_{medium}-k^2}}\rvert^2)I(\omega,k) g(\omega) dk d\omega \label{Eq:NF}\\
g(\omega)&=\frac{\frac{\Delta\omega}{2\pi}}{(\omega-\omega_0)^2+(\Delta\omega/2)^2} \label{Eq:linewidth}
\end{align}
where $\gamma^0_{ij}$ is the spontaneous decay rate for an isolated emitter and $g(\omega)$ is the lineshape of the transition with a linewidth of $\Delta\omega$. We follow the formulation laid out in  Chance et al. to obtain the distance dependence of $\gamma_{ij}$ of an isotropic emitter due to the modification of density of states by the surface in the near-field \cite{chance_molecular_1978} .
  
To obtain the power densities using Eqs.\,\ref{Eq:P10} and \ref{Eq:P20}, we model the induced absorption rate $W_{01}$ due to far-field pumping using the well-known expression for the stimulated rate \cite{yariv_quantum_1989} $W_{ij,external}=\frac{ \lambda^2 g(\omega) I_v \gamma^r_{ij}}{8\pi  n^2 \hbar \omega }$ where $\gamma^r_{ij}$ is the radiative spontaneous decay rate that couples to external pumping from the far-field, $I_v$ is the incident intensity of the external pumping field and $n$ is the index of the chalcogenide medium. The linewidth for the 0-1 and 2-1 transitions are assumed to be $2\times10^{11}\,\text{s}^{-1}$, comparable to typical laser gain mediums \cite{yariv_quantum_1989}.

The intrinsic efficiency of thermal extraction versus distance from the emitter is shown in Fig.\,\ref{fig:eff_real}(b). The maximum efficiency is small, around 0.33\% and decreases to zero beyond a few hundred nanometers. The total extracted intensity is defined as the integral of the power emitted by the 2-0 transition over all distances, $\int_{z_1}^{z_2}P_{20,net}dz=\int_{z_1}^{z_2}(dI_{20,net}/dz) dz=\Delta I_{20,net}$. We integrate from 10 nm onward until the intrinsic efficiency decreases to almost zero. Figure \ref{fig:eff_real}(c) shows the extracted power per unit area as a function of input power $I_v$. The extracted power increases linearly with the input power for low power inputs before saturating at higher powers, but the overall power extracted is orders of magnitude lower than the input power. The saturation is due to a high pump power which causes the population difference between the levels to tend to zero, leading to a maximum in absorption and thus the saturation behavior. A limiting case of Eq.\,\ref{Eq:P20} can be found for large $W_{01}$ as $\hbar(\omega_{20}-\omega_{10})W_{12}\gamma^r_{20} N_t /(3W_{12}+2(\gamma_{20}+\gamma_{12}))$. Integrating this limit over distance agrees with the saturation as plotted in Fig.\,\ref{fig:eff_real}(c).

\begin{figure}[h!]
\includegraphics[scale=0.5]{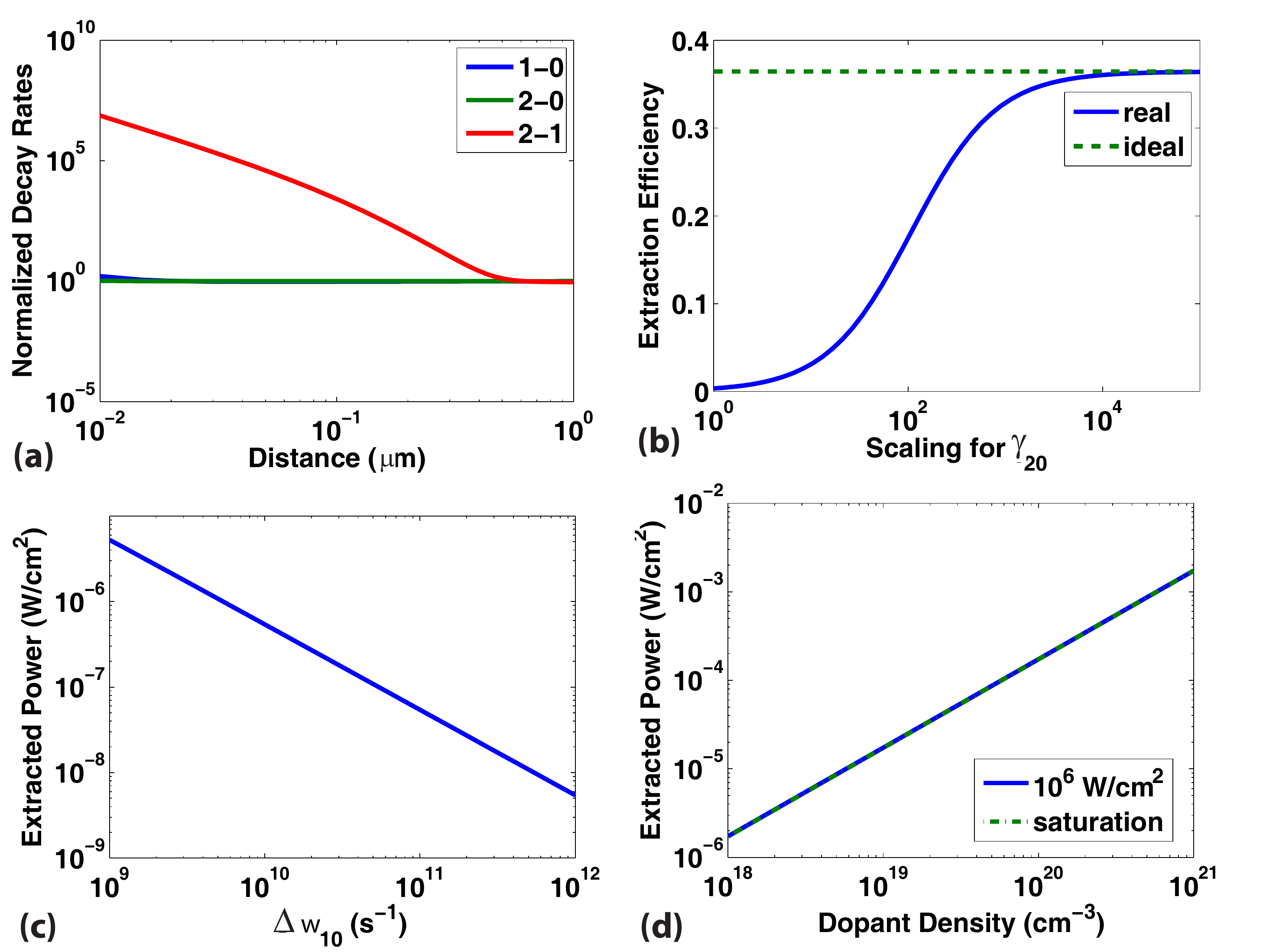}
\caption{\label{fig:gamma21} (a) Normalized spontaneous decay rates versus distance for three different transitions. The 2-1 transition is on resonance with the substrate dispersion and is enhanced greatly whereas the 0-1 and 2-0 transitions are not significantly affected by the presence of the substrate.  (b) Intrinsic extraction efficiency $\eta_{10}$ versus the scaling of the spontaneous rate $\gamma_{20}$ at $d=20$ nm. The blue line shows real behavior according to Eq.\,\ref{Eq:eff10}. Increasing $\gamma_{20}$ greatly enhances the efficiency so much so that around $10^4 \gamma_{20}$ the efficiency approaches the ideal limit of the system. (c) Integrated extracted power versus the linewidth of the 0-1 transition for a input power of $0.01\,\text{W/cm}^2$ in the linear regime with the same parameters as Fig.\,\ref{fig:eff_real}(b). Decreasing the linewidth increases the absorption efficiency and therefore the output power. (d) Integrated extracted power for different emitter densities in the saturation regime with an input power of $10^{6}\,\text{W/cm}^2$. The full description using Eq.\,\ref{Eq:P20} at high input power agrees with the saturation limit shown in the dashed green line. Increasing the emitter density increases the saturation extracted power by the same order of magnitude.}
\end{figure}

Figure \ref{fig:eff_real} shows that active thermal extraction is possible, but both the intrinsic efficiency and the total power extracted are very small for the chosen parameters. However, according to the limit of Eq.\,\ref{Eq:eff10}, the maximum efficiency should be around 35\%, much higher than in the example. To understand the reason for this difference, we examine Eq.\,\ref{Eq:eff10} in more detail. The maximum efficiency occurs when $\gamma_{20}$ and $W_{12}$ are large. We calculate the transition rates versus distance from the substrate in Fig.\,\ref{fig:gamma21}(a), and observe that the transition rates for 0-1 and 2-0 transitions are not affected by the presence of a surface as they are off-resonant. However, the decay rate for the 1-2 transition $\gamma_{12}$ is strongly enhanced as the emitter approaches the surface \cite{chance_molecular_1978,shimizu_surface-enhanced_2002,okamoto_surface-plasmon-enhanced_2004}. As a result, the near-field absorption rate is smaller by about two orders of magnitude compared to the decay rate even though both are enhanced by orders of magnitude due to the increase in the optical density of states in the near-field. Physically, this calculation indicates that as electrons are excited from energy level 1 to 2, they immediately decay back to level 1 at the rate $\gamma_{12}$.

The reason for this cycling is that the thermal near-field energy density is not sufficient to allow near-field absorption to dominate over near-field spontaneous decay. Archambault et al.\,\cite{archambault_quantum_2010} also highlight the need for some minimum energy density for stimulated emission to dominate spontaneous decay. Unlike the case for stimulated emission of surface plasmons with external pumping such as in Ref.\,\cite{noginov_demonstration_2009} where the external laser field intensities can be tuned, here the thermal energy density is restricted to that for a blackbody. Thus, the spontaneous decay rate will always dominate over near-field absorption for realistic values of near-field energy density. On the other hand, Fig.\,\ref{fig:eff_real}(a) also shows that while a resonantly enhanced $\gamma_{12}$ offsets the enhanced absorption $W_{12}$, the extraction efficiency $\eta_{10}$ still requires a large value of $W_{12}$. Beyond a emitter-substrate distance of about 100 nm, the extraction efficiency in Fig.\,\ref{fig:eff_real}(a) drops significantly as a result of the low near-field energy density, although the ratio $W_{12}/\gamma_{12}$ remains in the same order of magnitude up to a distance of 1 $\mu$m.

Therefore, to break the cycling between levels 1 and 2, it is essential that the strongly radiative decay rate from 2-0 ($\gamma_{20}$) is comparable to the decay rate $\gamma_{12}$ in the near-field. Figure \ref{fig:gamma21}(b) shows that the efficiency is boosted to almost the ideal limit at short distances if $\gamma_{20}$ is increased substantially. In Eq.\,\ref{Eq:eff10}, if we increase $\gamma_{20}$ to be more comparable to $\gamma_{12}$ in the near-field, then the ratio of $\gamma^r_{20}/\gamma_{20}$ begins to dominate in the expression, increasing the extraction efficiency towards the ideal limit discussed earlier. 

The factors discussed above affect the intrinsic efficiency, but the total extracted power also depends on the input power $W_{01}$ and the emitter density $N_t$. We now examine the role of these parameters. Firstly, the absorption of the pump power $W_{01}$ depends on the linewidth of the 0-1 transition, and decreasing the linewidth increases $W_{01}$ in Eq.\,\ref{Eq:P20} due to the increased concentration of input power in a given bandwidth for each emitter. Figure \ref{fig:gamma21}(c) shows that decreasing the linewidth of the 0-1 transition increases the integrated extracted power as predicted. Secondly, the total dopant density $N_t$ also affects the extracted power. As discussed earlier, the saturation limit at higher incident powers is proportional to the dopant density, and therefore the dopant density must increase to increase the saturation limit. Figure \ref{fig:gamma21}(d) shows the integrated extracted power at a high input power as a function of emitter density. The extracted power scales accordingly with the dopant density and agrees well with the prediction from the expression for the saturation limit. Thus, increasing the dopant density shifts the saturation limit in Fig.\,\ref{fig:eff_real}(c) up proportionally.

\begin{figure}
\includegraphics[scale=0.55]{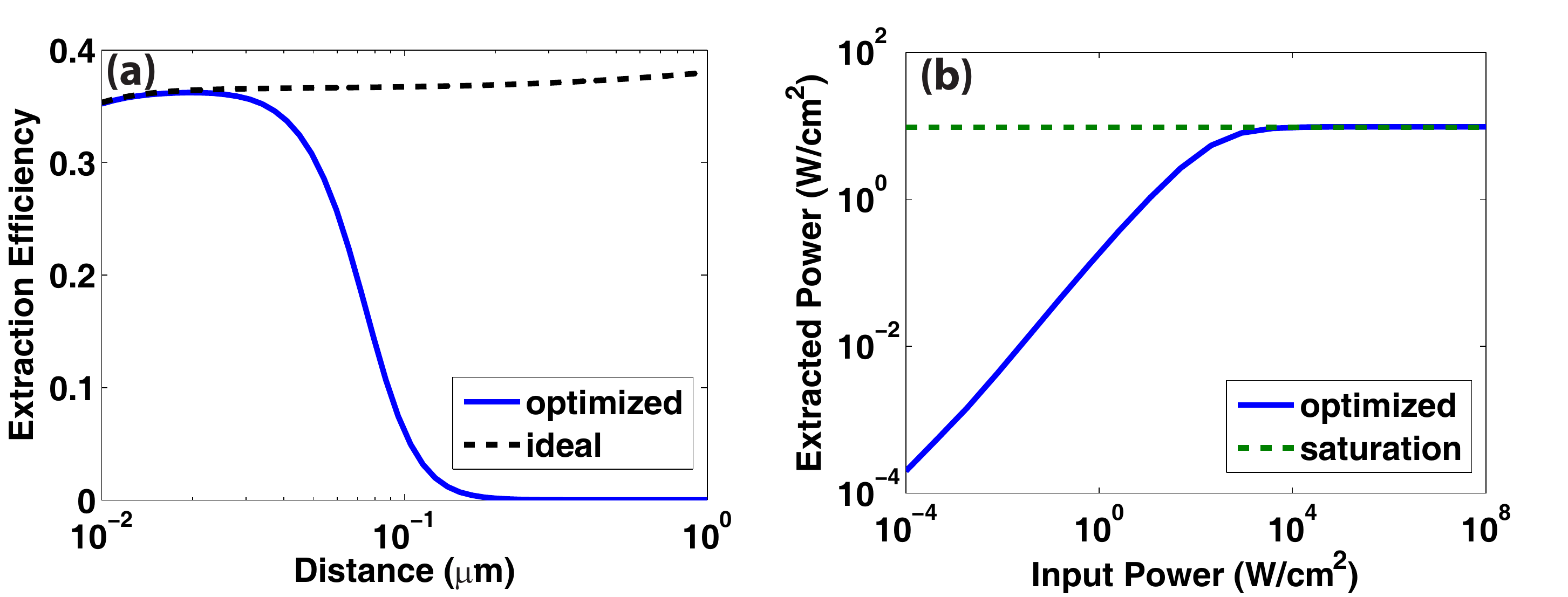}
\caption{\label{fig:eff_ideal} (a) Intrinsic extraction efficiency versus emitter-substrate distance for an optimized system. The extraction efficiency follows the ideal limit for small distances before decreasing due to a decreasing $W_{12}$. The extraction efficiency is much improved compared to Fig.\,\ref{fig:eff_real}(b). (b) Integrated power extracted of the optimized system with emitters uniformly distributed from 10 nm onward from the substrate surface. A reduced 0-1 transition linewidth of $\Delta\omega_{10}=2\times10^9\,\text{s}^{-1}$ and a higher emitter density $N_t=10^{21}\,\text{cm}^{-3}$ lead to much higher saturation limit shown in the dashed line.} 
\end{figure}

Using this understanding, we now recalculate the efficiency and extracted power for an optimized gain medium with the spontaneous rate for the 2-0 transition increased to $1.37\times10^8\,\text{s}^{-1}$,  $\Delta \omega_{10}=2\times10^{9}\,\text{s}^{-1}$ and $N_t=10^{21}\,\text{cm}^{-3}$. Figure \ref{fig:eff_ideal}(a) shows that the intrinsic extraction efficiency is much higher than in Fig.\,\ref{fig:eff_real}(b) and almost near the ideal limit for small emitter-substrate distances. The decrease of efficiency at larger emitter-substrate distances is due to a decrease in near-field coupling. Figure \ref{fig:eff_ideal}(b) shows a much-increased integrated extracted power at each given input power compared to Fig.\,\ref{fig:eff_real}(c). The saturation limit derived earlier also agrees with the full calculation at higher input powers. This limit is independent of the input power and is proportional to the dopant density.

This calculation shows that the active thermal extraction scheme has potential to efficiently extract a significant amount of near-field thermal radiative energy. The key to realizing this potential is to identify an appropriate gain medium with surface resonance and a emitter with matching transitions in the mid-infrared wavelength range where photons are thermally populated at typical temperatures. Decreasing the linewidth (such as using a cavity) or increasing the 0-1 spontaneous decay rate helps to decrease the input power necessary for extracting a given power, but a high dopant density is still required to increase the saturation limit. Cerium doped crystals can potentially be a candidate as they have a $4f^05d^1\rightarrow4f^15d^0$ transition with a short lifetime of around 40 ns \cite{scharmer_efficient_1982}, ideal for the 2-0 transition proposed here, as well as a mid-infrared transition of 4.5 $\mu$m \cite{scharmer_efficient_1982} for the near-field absorption. Radium is a very good candidate with parameters such as transition wavelengths and decays rates comparable to our optimized model. It has been recently demonstrated to interact with black body radiation in a laser trapping experiment \cite{guest_laser_2007}.

Our work shares some similarities with laser cooling of solids \cite{sheik-bahae_optical_2007,seletskiy_laser_2010,zhang_laser_2013,khurgin_surface_2007} and active schemes in plasmonics \cite{cuerda_theory_2015}, photonic crystals \cite{bermel_active_2006}, and metamaterials \cite{wuestner_overcoming_2010,ni_loss-compensated_2011} but differs in a number of important ways. First, laser cooling directly extracts phonons, while our scheme extracts surface phonon polaritons. Therefore, our scheme has potential to be much more efficient than laser cooling because of the significantly higher energy of surface phonon polaritons than phonons. Also, laser cooling requires the medium to be cooled to possess very specific energy levels, whereas our scheme only requires that the medium possess a surface resonance. Second, active schemes in plasmonics have been used to realize spasers and to compensate loss. One main distinction between these schemes and our approach is that here one transition is pumped by a near-field, incoherent thermal radiative source rather than a coherent pump. As a result, our approach does not lead to any form of stimulated emission or coherent single mode emission. Additionally, the thermal energy density is not sufficient to cause the imaginary part of permittivity of the gain medium to become positive; our medium is actually absorptive under all conditions.

In conclusion, we have numerically demonstrated an active thermal extraction scheme that allows bound surface waves to be converted from evanescent to propagating waves. Our work exploits the monochromatic nature of near-field radiation to drive a transition in a gain medium simultaneously with an external pump. This demonstrates shows the high potential for manipulating thermal radiation using active processes rather than the traditional passive approaches.

\begin{acknowledgments}
This work is part of the 'Light-Material Interactions in Energy
Conversion' Energy Frontier Research Center funded by the U.S. Department
of Energy, Office of Science, Office of Basic Energy Sciences under Award
Number DE-SC0001293. D.D. gratefully acknowledges the support by the
Agency for Science, Technology and Research (Singapore). A. J. M. acknowledges the support of the Northrop Grumman Corporation.
\end{acknowledgments}


%

\end{document}